\documentclass[12pt]{imsart}

\usepackage{enumitem}
\usepackage[utf8]{inputenc}
\usepackage[english]{babel}
\usepackage[margin=1in]{geometry}
\usepackage[colorlinks,citecolor=blue,urlcolor=blue]{hyperref}

\usepackage{bm}
\usepackage{graphicx}
\usepackage{amssymb,amsmath,amsthm,mathrsfs}
\usepackage{epstopdf}
\usepackage{algorithmic}
\usepackage{xcolor}
\usepackage{subfigure}
\usepackage{multirow}
\usepackage{titlesec,textcase}
\usepackage{longtable}
\usepackage{rotating}
\usepackage{booktabs}
\usepackage{siunitx} 
\usepackage{float}

\newtheorem{Theorem}{Theorem}[section]
\newtheorem{Lemma}[Theorem]{Lemma}

\newtheorem{Proposition}[Theorem]{Proposition}

\newtheorem{Definition}[Theorem]{Definition}

\theoremstyle{definition}

\newcommand{\diag}{\mathrm{diag}}

\RequirePackage[normalem]{ulem} 
  
\def\onebm{{\bm{1}}}

\DeclareMathOperator{\tr}{tr}
\definecolor{rp}{RGB}{83,54,106}

\def\boxit#1{\vbox{\hrule\hbox{\vrule\kern6pt\vbox{\kern6pt#1\kern6pt}\kern6pt\vrule}\hrule}}

\begin{document}
\begin{frontmatter}
\title{Hypothesis testing for the dimension of random geometric graph}

\runtitle{Test the dimension of RGG }
\runauthor{ M. Yuan and F. Yu }
 
\begin{aug}
\author[B]{\fnms{Mingao} \snm{Yuan}\thanks{Mingao Yuan is grateful for the generous support provided by the Startup Fund of The University of Texas at El Paso.}\ead[label=e2]{myuan2@utep.edu}},
\author[B]{\fnms{Feng} \snm{Yu}\ead[label=e3]{fyu@utep.edu}}

\address[B]{Department of Mathematical Sciences,
The University of Texas at El Paso, El Paso, TX, USA\\
\printead{e2}; \printead{e3}}

 


\end{aug}
 
\begin{abstract}
Random geometric graphs (RGGs) offer a powerful tool for analyzing the geometric and  dependence structures in real-world networks. For example, it has been observed that RGGs are a good model for protein-protein interaction networks. In RGGs, nodes are randomly distributed over an $m$-dimensional metric space, and edges connect the nodes if and only if their distance is less than some threshold. When fitting RGGs to real-world networks, the first step is probably to input or estimate the dimension $m$. However, it is not clear whether the prespecified dimension is equal to the true dimension. In this paper, we investigate this problem using hypothesis testing. Under the null hypothesis, the dimension is equal to a specific value, while the alternative hypothesis asserts the dimension is not equal to that value. We propose the first statistical test. Under the null hypothesis, the proposed test statistic converges in law to the standard normal distribution, and under the alternative hypothesis, the test statistic is unbounded in probability. We derive the asymptotic distribution by leveraging the asymptotic theory of degenerate U-statistics with kernel function dependent on the number of nodes. This approach differs significantly from prevailing methods used in network hypothesis testing problems. Moreover, we also propose an efficient approach to compute the test statistic based on the adjacency matrix.
 Simulation studies show that the proposed test performs well. We also
apply the proposed test to multiple real-world networks to test their dimensions.
\end{abstract}

\begin{keyword}[class=MSC2020]
\kwd[]{62A09}
\kwd[;  ]{62E20}
\end{keyword}

\begin{keyword}
\kwd{hypothesis test}
\kwd{random geometric graph}
\kwd{dimension}
\end{keyword}

\end{frontmatter}

\section{Introduction}
Graph is a mathematical structure that models pairwise relations between objects.
A graph or network consists of a set of nodes or vertexes connected by edges. 
It is a powerful mathematical model that has been widely used to solve many real-world problems \cite{A17,CGL16,N03,CC12,TPLBMTM22,CLHXHMR23,K09,BM08,SBL13}. For example, graph is used to analyze brain functional magnetic resonance imaging and understand normal brain function and brain disorders \cite{SBL13}. Biological networks are used to identify important molecules and identify topological features of molecular interactions \cite{CBL16}.
In sociology, network provides a powerful approach to study dependence structures among social units \cite{BM08}.

Graphs or networks are high-dimensional data structures. In order to efficiently extract information, we need to assume some structure on networks. To this end, a large number of random graph models have been introduced \cite{DC23,SHL11}. The most popular and simplest one is the Erd\H{o}s-R\'{e}nyi model, where nodes are randomly and independently connected with the same probability. Many real-world networks exhibit  geometry and dependence structure \cite{DC23,HRP08,GMPS23}. Random Geometric Graphs, which are variations of the classic Erd\H{o}s-R\'{e}nyi random graph,  have been introduced to model such networks \cite{DC23,HRP08,GMPS23}. 
Each node in RGGs is randomly assigned a position in an $m$-dimensional metric space, and two nodes are connected by an edge if and only if their distance is smaller than some prespecified threshold \cite{DC23}. Edges in RGGs are therefore correlated and edge existence is related to the distance between nodes.  Geometry plays an important role in RGGs, resulting in networks that have rich dependence structures and exhibit desirable real-world network features.

 RGGs serve as a model for dependence structure and geometry in a variety of real networks \cite{DC23,HRP08,GMPS23,BJ23,SSH18,ZLKG14}. Their 
applications range from network user profiling to protein-protein interactions in biology. For example, RGGs effectively represent  protein–protein
interaction networks \cite{HRP08}. It plays an important role in  
discovering new biological features and understanding the interplay between network structure
and biological function \cite{HRP08,KRHP09}.  RGGs are also used to
understand  the spatial distribution and propagation of opinions in social
dynamics \cite{ZLKG14}.

In the application of RGGs, one of the fundamental problems is that the underlying dimension $m$ is unknown. One needs to estimate the dimension or input a prior dimension.  For example, \cite{HRP08} fits a RGG with $m=2,3,4$ to protein-protein interaction networks; \cite{ADL23} provides several consistent estimators of the dimension of a RGG. However, it is not clear whether the prespecified or estimated dimension is significantly different from the true dimension. In this paper, we address this problem by using hypothesis testing. Under the null hypothesis, the true dimension is equal to the specified value; under the alternative hypothesis, the true value is not equal to that specified value. We introduce the first statistical test for the hypotheses. We prove that, under the null hypothesis, the proposed test statistic converges in distribution to the standard normal distribution. The asymptotic distribution of the test statistic is established using degenerate U-statistic theory, where the kernel function is dependent on the network size. This approach contrasts with standard methods in network hypothesis testing problems. A theoretical investigation of the  power is also performed. The theoretical results are supported by simulations and data examples.

\vskip 5mm

\noindent
{\bf Notation:} We adopt the  Bachmann–Landau notation throughout this paper. Given two sequences  $a_n$  and $b_n$, denote $a_n=\Theta(b_n)$ if $c_1\leq \left|\frac{a_n}{b_n}\right |\leq c_2$ for some positive constants $c_1,c_2$ and large $n$. Denote  $a_n=\omega(b_n)$ if $\lim_{n\rightarrow\infty}\left|\frac{a_n}{b_n}\right |=\infty$. Denote $a_n=O(b_n)$ if $\left|\frac{a_n}{b_n}\right |\leq c$ for some positive constant $c$ and large $n$. Denote $a_n=o(b_n)$ if $\lim_{n\rightarrow\infty}\left|\frac{a_n}{b_n}\right |=0$. Let $N(0,1)$ be the standard normal distribution. For a sequence of random variables $X_n$, $X_n\Rightarrow N(0,1)$ represents $X_n$ converges in distribution to $N(0,1)$ as $n$ goes to infinity. $\mathbb{E}[X_n]$ and $Var(X_n)$ denote the expectation and variance of a random variable $X_n$ respectively. $\mathbb{P}(E)$ denote the probability of an event $E$. Given a finite set $E$, $|E|$ represents the number of elements in $E$. Given positive integer $t$, $\sum_{i_1\neq i_2\neq\dots\neq i_t}$ means summation over all integers $i_1,i_2,\dots,i_t$ in $[n]=\{1,2,\dots,n\}$ such that $|\{i_1,i_2,\dots,i_t\}|=t$. $\sum_{i_1< i_2<\dots< i_t}$ means summation over all integers $i_1,i_2,\dots,i_t$ in $[n]$ such that $i_1<i_2<\dots<i_t$.

\section{The hypotheses and the proposed test}

A graph consists of a set of nodes and a set of edges. An edge joins two nodes and models the interaction between them. A graph is typically represented as an adjacency matrix $A$. Given a graph on $n$ nodes, $A_{ij}=1$ indicates the presence of an edge between node $i$ and node $j$; $A_{ij}=0$, otherwise. Throughout this paper, we assume the graph is undirected and there is no self-loop. That is, the adjacency matrix $A$ is symmetric, and the diagonal elements of $A$ are zero.

 Random graph is a probability distribution over graphs, in which edges between  nodes are ramdomly placed. The most popular random graph is the Erd\H{o}s-R\'{e}nyi 
random graph, where nodes are randomly and independently connected with the same probability. Random Geometric Graph  is an undirected graph constructed by randomly placing $n$ nodes in some metric space and connecting two nodes by an edge if and only if their distance is less than a threshold. Random Geometric Graph is a generalization of the Erd\H{o}s-R\'{e}nyi 
random graph. It has been widely used to model the intrinsic geometry and dependence structure of real-world networks \cite{HM09,HM12,HRP08,DC23,GMPS23}.  
Random Geometric Graph is a versatile
graph family, each offering different ways to model spatial relationships and network structures.  In this paper, we adopt the following Random Geometric Graph \cite{DC23,GMPS23}.

\begin{Definition}\label{def1}
Let $r_n\in[0,0.5]$ be a real number, and $m$ be a positive integer. Given i.i.d. uniform random variables $X_1,X_2,\dots,X_n$ on the unit square $[0,1]^m$, the Random Geometric Graph (RGG) $\mathcal{G}_{n}(m,r_n)$ is defined as
\[A_{ij}=I[d(X_i,X_j)\leq r_n],\]
 where $A_{ii}=0$, $X_i=(X_{i1},X_{i2},\dots,X_{im})$ and 
 \begin{eqnarray}\label{distance}
 d(X_i,X_j)=\max_{1\leq k\leq m}\Big\{\min\{|X_{ik}-X_{jk}|,1-|X_{ik}-X_{jk}|\}\Big\}.
 \end{eqnarray}
\end{Definition}
The unit square $[0,1]^m$ is a common choice of the metric space for the nodes in RGG  \cite{HRP08,GRK05,Z15}.  The distance (\ref{distance}) is widely used in defining RGG \cite{BKL19, BB24}.  The unit square $[0,1]^m$ equipped with the distance (\ref{distance}) is  homeomorphic to the $m$-dimensional torus. In $\mathcal{G}_{n}(m, r_n)$, each node is  uniformly and independently placed on $[0,1]^m$, and two nodes are joined by an edge if their distance is less than or equal to $r_n$.
 The parameter $r_n$ regulates the level of sparsity of the graph. The RGG with $r_n\geq0.5$ produces a complete graph; that is, every pair of nodes is connected by an edge. When $r_n=0$, all the nodes are isolated.  The RGG $\mathcal{G}_{n}(m, r_n)$ is a good model for networks with geometry and dependence structure  \cite{DC23,GMPS23,HRP08,KRHP09}, since the edges in it are correlated and dependent on the distance.

When fitting RGG $\mathcal{G}_{n}(m,r_n)$ to real-world networks, the first step is to input or estimate the dimension $m$ \cite{HRP08, ADL23}.   However, it is not clear whether the dimension $m$ is equal to the prespecified or estimated  value. In this paper, we address this problem by hypothesis testing.  
Given a network $A\sim \mathcal{G}_{n}(m,r_n)$, we are interested in testing the following hypotheses
\begin{equation}\label{hypo}
H_0:  m=m_0,\hskip 1cm H_1: m\neq m_0,
\end{equation}
where $m_0$ is a prespecified positive integer.

In this paper, we propose the first statistical test for (\ref{hypo}). To this end, 
we define a pivotal statistic $D_n$ as follows
\begin{equation}\label{dn}
D_n=\sum_{i\neq j\neq k}A_{ij}A_{jk}A_{ki}-\left(\frac{3}{4}\right)^{m_0}\sum_{i\neq j\neq k}A_{ij}A_{ik}.
\end{equation}
Next, we find the order of $D_n$ of the RGG $\mathcal{G}_{n}(m,r_n)$.

\begin{Proposition}\label{prop}
Suppose $A\sim \mathcal{G}_{n}(m,r_n)$ with $r_n=o(1)$ and $nr_n^{m}=\omega(1)$. Then
\begin{eqnarray}\label{dnorder}
D_n=n(n-1)(n-2)r_n^{2m}\left[3^m-\left(\frac{3}{4}\right)^{m_0}4^{m}\right]+O_P\left(\frac{n^3r_n^{2m}}{n\sqrt{r_n^m}}\right).
\end{eqnarray}
\end{Proposition}
Based on Proposition \ref{prop}, $D_n$ is equal to
\[D_n=O_P\left(\frac{n^3r_n^{2m}}{n\sqrt{r_n^m}}\right),\]
if  the null hypothesis $H_0$ holds.   
 Under the alternative hypothesis $H_1$, $D_n$ is equal to
\[D_n=n^3r_n^{2m}+O_P\left(\frac{n^3r_n^{2m}}{n\sqrt{r_n^m}}\right).\]
The quantity $D_n$ has different orders under the null hypothesis $H_0$ and the alternative hypothesis $H_1$. This implies that suitably scaled $D_n$ has high potential to be a powerful test statistic for (\ref{hypo}).

\begin{Theorem}\label{thm1}
Suppose $A\sim \mathcal{G}_{n}(m,r_n)$ with $r_n=o(1)$ and $nr_n^{m}=\omega(1)$. Under the null hypothesis $H_0$, we have
\begin{equation*}
    \frac{\sqrt{2}D_n}{n^2\hat{\sigma}_{n2}}\Rightarrow N(0,1).
    \end{equation*}
 Here, $D_n$ is defined in (\ref{dn}), and
\begin{eqnarray*} \nonumber
\hat{\sigma}_{n2}^2
&=&\left[36-24\left(\frac{3}{4}\right)^{m_0}\right]S_1+\left[16\left(\frac{3}{4}\right)^{2m_0}-48\left(\frac{3}{4}\right)^{m_0}\right]S_2\\
&&+8\left(\frac{3}{4}\right)^{2m_0}S_3+4\left(\frac{3}{4}\right)^{2m_0}S_4+8\left(\frac{3}{4}\right)^{2m_0}S_5,
\end{eqnarray*}
where
\begin{eqnarray*}  
S_1&=&\frac{1}{n^4}\sum_{i\neq j\neq k\neq l}A_{ij}A_{jk}A_{kl}A_{li}A_{ik},\\
S_2&=&\frac{1}{n^4}\sum_{i\neq j\neq k\neq l}A_{ij}A_{jk}A_{ki}A_{il},\\
S_3&=&\frac{1}{n^4}\sum_{i\neq j\neq k\neq l}A_{ij}A_{ik}A_{il},\\
S_4&=&\frac{1}{n^4}\sum_{i\neq j\neq k\neq l}A_{ij}A_{jk}A_{kl}A_{li},\\
S_5&=&\frac{1}{n^4}\sum_{i\neq j\neq k\neq l}A_{ij}A_{jk}A_{kl}.
\end{eqnarray*}
\end{Theorem}

The assumptions of Theorem \ref{thm1} are weak. The condition $r_n=o(1)$ means that the network is sparse; that is, the
total number of edges in the network is $o(n^2)$. This assumption is not restrictive, because most real-world networks are sparse \cite{A17}. The condition $nr_n^{m}=\omega(1)$ means that  the total number of the network tends to infinity. This assumption is common in the theoretical analysis of RGGs \cite{GMPS23}.

The proof of Theorem \ref{thm1}  is based on representing $T_n$ as the sum of a degenerate U-statistic and a remainder. Unlike standard degenerate U-statistics, the kernel function of the U-statistic in our paper  is $n$-dependent. Standard asymptotic theory for degenerate U-statistics is not applicable. We instead employ the asymptotic theory for degenerate U-statistics with a kernel that varies with $n$ in \cite{FL96}. In this scenario, the asymptotic distribution of the degenerate U-statistic, as shown in Theorem \ref{thm1}, is the standard normal. Unlike the typical constant kernel scenario, this result yields an asymptotic distribution that is not a sum of independent chi-square distributions.  Our method deviates from existing approaches in the context of network hypothesis testing problems \cite{L16,GL17,JKL18,YLFS19}.

 According to Theorem \ref{thm1}, $ \frac{\sqrt{2}D_n}{n^2\hat{\sigma}_{n2}}$ converges in law to the standard normal distribution, as the number of nodes $n$ goes to infinity. Given Type I error $\alpha$, we define \textit{the dimension test} as follows:
\begin{equation}\label{rejectionregion}
    \text{Reject $H_0$,\  \ if $\left| \frac{\sqrt{2}D_n}{n^2\hat{\sigma}_{n2}}\right|\geq Z_{\frac{\alpha}{2}}$, }
\end{equation}
where $Z_{\frac{\alpha}{2}}$ is the $100(1-\frac{\alpha}{2})\%$ quantile of the standard normal distribution. Based on Theorem \ref{thm1}, the Type I error of \textit{the dimension test} is asymptotically equal to $\alpha$.

Subsequently, we examine the power of the proposed dimension test.
\begin{Theorem}\label{thm2}
Suppose $A\sim \mathcal{G}_{n}(m,r_n)$ with $r_n=o(1)$ and $nr_n^{m}=\omega(1)$. Under the alternative hypothesis $H_1$, we have
\begin{equation}\label{powerorder}
    \frac{\sqrt{2}D_n}{n^2\hat{\sigma}_{n2}}=\Theta\big(n\sqrt{r_n^m}\big)+O_P(1).
    \end{equation}
\end{Theorem}

Since $nr_n^{m}=\omega(1)$ by assumption, then $n\sqrt{r_n^m}=\sqrt{n}\sqrt{nr_n^m}=\omega(1)$. Under the alternative hypothesis $H_1$, the test statistic $\frac{\sqrt{2}D_n}{n^2\hat{\sigma}_{n2}}$ goes to infinity. Therefore, the power of the proposed dimension test approaches one as
$n$ goes to infinity.

\noindent\textbf{Computational Efficiency and Implementation.} The quantities $D_n, S_1, S_2, S_3, S_4, S_5$ involve summations over three or four indices. A direct evaluation through nested for-loops is highly inefficient in most programming languages and is prohibitively expensive for large $n$. In order to address this issue, we provide equivalent matrix-form expressions of these quantities as follows:
\begin{align*}
    D_n = & \tr(A^3)+2\tr(A) -  \left(\frac{3}{4}\right)^{m_0}\Big(\onebm^\top (A^2-A)\onebm+2\tr(A)\Big),\\
    S_1 = & \onebm^\top[A^2\odot A^2 \odot A-A^2\odot A]\onebm, \\
    S_2 = & \diag(A^3)^\top A\onebm-2\tr(A^3), \\
    S_3 = & \onebm^\top(A\onebm)^{\circ 3}+\onebm^\top (2A-3A^2)\onebm, \\
    S_4 = & \tr(A^4)+\onebm^\top (A-2A^2)\onebm, \\
    S_5  = & \onebm^{\top}(A^3-2A^2+A)\onebm-\tr(A^3).
\end{align*}
Here, $\mathbf{1}=(1,1,\dots,1)^\top\in\mathbb{R}^{n}$ denotes an $n$-dimensional vector of ones. For a matrix $M\in\mathbb{R}^{n\times n}$, $M^\top\in\mathbb{R}^{n\times n}$ its transpose and $\diag(M)\in\mathbb{R}^n$ denotes the vector formed by the diagonal entries of $M$. 
The symbol $\odot$ denotes Hadamard product (or element-wise product) of two matrices, i.e. $[M_1\odot M_2]_{i,j}=[M_1]_{ij}[M_2]_{ij}$. We use $M^{\circ k}$ to denote the Hadamard power, that is, the elementwise product of $k$ copies of $M$. 

The matrix expressions above can be derived from Theorem 3.2 in \cite{YY25} using the fact that $A^{\circ2}=A$. 


\section{Simulations and real data applications}

In this section, simulations and real-data applications are provided to support the theoretical results.

\subsection{Simulation studies}

We conduct simulation studies to evaluate the performance of the proposed dimension test. Throughout, the Type I error is set to be $\alpha=0.05$, and the networks are generated from the RGG $\mathcal{G}_{n}(m,r_n)$ in Definition \ref{def1}.    To simulate Type I error, we generate 1000 networks from $\mathcal{G}_{n}(m,r_n)$ under the null hypothesis,   perform the proposed dimension test  in (\ref{rejectionregion}) for each network, and record whether the null hypothesis is rejected. The empirical Type I error is equal to the proportion of rejection among the 1000 tests. The empirical power calculation remains the same, the only difference being that networks are generated according to the alternative hypothesis.

In the first simulation, we consider the following hypotheses:
\[H_0: m=1,\hskip 1cm H_1: m\neq 1.\]
Under the null hypothesis, the dimension is equal to one.  
To simulate the powers,
 we consider $m=2$ and $m=3$ in the alternative hypothesis. The networks are generated from $\mathcal{G}_{n}(m,r_n)$ with $r_n\in\{0.09, 0.10, 0.11\}$ and $n\in\{70, 100,130\}$. Table \ref{simum1} summarizes the results. The column with $m=1$ records the empirical Type I errors, and the columns with $m=2$ and $m=3$ show the empirical powers of the proposed test.


\begin{table}[!ht]
\centering
\caption{Empirical Type I errors and powers with $H_0: m=1$.}
\label{simum1}
\begin{tabular}{@{}cc *{3}{S[table-format=1.3]}@{}}
\toprule
\multirow{2}{*}{$r_n$} & \multirow{2}{*}{$n$} & \multicolumn{3}{c}{$m$} \\ 
\cmidrule(l){3-5}
 & & {1} & {2} & {3} \\
\midrule
\multirow{3}{*}{0.09} 
 & 70  & 0.049 & 0.920 & 0.645 \\ 
 & 100 & 0.046 & 0.990 & 0.834 \\ 
 & 130 & 0.040 & 1.000 & 0.952 \\ 
\midrule
\multirow{3}{*}{0.10} 
 & 70  & 0.047 & 0.953 & 0.742 \\ 
 & 100 & 0.042 & 0.999 & 0.903 \\ 
 & 130 & 0.046 & 1.000 & 0.985 \\ 
\midrule
\multirow{3}{*}{0.11} 
 & 70  & 0.045 & 0.981 & 0.805 \\ 
 & 100 & 0.041 & 1.000 & 0.973 \\ 
 & 130 & 0.043 & 1.000 & 0.997 \\ 
\bottomrule
\end{tabular}
\end{table}

The empirical Type I errors are close to the nominal level 0.05. This indicates that the distribution of the test statistic  is well approximated by the standard normal distribution, even for smaller network sizes. The proposed dimension test has high power. As $n$ or $r_n$ increases, the power approaches one. When $m$ changes from 2 to 3, $n\sqrt{r_n^m}$ decreases, resulting in a lower power for the dimension test. These phenomena are consistent with our theoretical results in Theorem \ref{thm1} 
and Theorem \ref{thm2}.

In the second simulation, we consider the following hypotheses:
\[H_0: m=2,\hskip 1cm H_1: m\neq 2.\]
Under the null hypothesis, the dimension equals two.  We simulate the powers with $m = 1$ and $m = 3$ in the alternative hypothesis. The networks are sampled from $\mathcal{G}_{n}(m,r_n)$ with $r_n\in\{ 0.27,0.29,0.31\}$ and $n\in\{40,50,60\}$. Table \ref{simum2} displays the results. The Type I errors are presented within the column corresponding to $m=2$, and the powers are shown in the columns with $m=1$ and $m=3$.


\begin{table}[!ht]
\centering
\caption{Empirical Type I errors and powers with $H_0: m=2$.}
\label{simum2}
\begin{tabular}{@{}cc *{3}{S[table-format=1.3]}@{}}
\toprule
\multirow{2}{*}{$r_n$} & \multirow{2}{*}{$n$} & \multicolumn{3}{c}{$m$} \\ 
\cmidrule(l){3-5}
 & & {2} & {1} & {3} \\
\midrule
\multirow{3}{*}{0.27} 
 & 40 & 0.064 & 1.000 & 0.927 \\ 
 & 50 & 0.057 & 1.000 & 0.980 \\ 
 & 60 & 0.044 & 1.000 & 0.999 \\ 
\midrule
\multirow{3}{*}{0.29} 
 & 40 & 0.057 & 1.000 & 0.951 \\ 
 & 50 & 0.048 & 1.000 & 0.992 \\ 
 & 60 & 0.047 & 1.000 & 1.000 \\ 
\midrule
\multirow{3}{*}{0.31} 
 & 40 & 0.054 & 1.000 & 0.976 \\ 
 & 50 & 0.043 & 1.000 & 0.993 \\ 
 & 60 & 0.044 & 1.000 & 0.999 \\ 
\bottomrule
\end{tabular}
\end{table}

The empirical Type I errors are in line with the expected 0.05 level. This indicates that the distribution of the test statistic is close to the standard normal distribution, even for smaller network sizes. The proposed dimension test has high power. Especially, the power for $m=1$ is equal to one under the this simulation setting.  When $n$ or $r_n$ increases, the power gets closer to one.  These findings align with our theoretical predictions in Theorem \ref{thm1} 
and Theorem \ref{thm2}.


\subsection{Real data applications}

We apply the proposed dimension test to multiple real-world networks available in \cite{HDATA} to infer their dimensions. These networks include four cheminformatics networks, one brain network and one tortoise social network.   In a cheminformatics network,  nodes are compounds, and edges are defined as a pairwise relationship such as a 2D fingerprint similarity value \cite{SP22}. In the brain network,  nodes are brain regions, and edges represent fiber tracts that connect one node to another \cite{HDATA}. In the tortoise social network, an edge between two tortoises represents sharing the burrow \cite{SET16}. Table \ref{realdata} lists the networks and their number of nodes and edge densities.  The first four are cheminformatics networks, the fifth one is the brain network, and the last one is the tortoise social network.

We test the null hypothesis that the dimension of each network is equal to a prespecified value $m_0$; that is, $H_0: m=m_0$. Many networks have been observed to have a small dimension \cite{HRP08}. We therefore consider $m_0\in\{1,2,3,4,5\}$. Then we perform the proposed dimension test defined in (\ref{rejectionregion}) and calculate its p-value.  Table \ref{realdata} lists the results. The highlighted are p-values greater than Type I error $\alpha=0.05$. At significance level $\alpha=0.05$, we fail to reject the null hypothesis $H_0: m=2$ for the network `ENZYMES-g147', and we reject the null hypothesis with $m_0\neq 2$. Therefore, we conclude that the network `ENZYMES-g147' has dimension $m_0=2$. Similarly, the networks `ENZYMES-g196' and `macaque-rhesus-brain-2' have dimension $m_0=3$, the network `ENZYMES-g532' has dimension $m_0=5$, and `reptilia-tortoise-network-bsv' has dimension $m_0=4$. The networks do not have the same dimension. This highlights the importance of testing the dimension of a network before fitting RGGs to it.


\begin{table}[h!]
\centering
\caption{P-values for testing the null hypothesis $H_0: m = m_0$.}
\label{realdata}
\begin{tabular}{@{}lccccccc@{}}
\toprule
\textbf{Network} & \textbf{$n$} & \textbf{Density} & \textbf{$m_0=1$} & \textbf{$m_0=2$} & \textbf{$m_0=3$} & \textbf{$m_0=4$} & \textbf{$m_0=5$} \\ 
\midrule
ENZYMES-g147 & 40 & 0.210 & 0 & \textbf{0.696} & 0 & 0 & 0 \\
ENZYMES-g196 & 50 & 0.138 & 0 & 0 & \textbf{0.653} & 0 & 0 \\
ENZYMES-g209 & 57 & 0.124 & 0 & 0.022 & 0 & 0 & 0 \\
ENZYMES-g532 & 74 & 0.085 & 0 & 0 & 0 & 0.002 & \textbf{0.140} \\
macaque-rhesus-brain-2 & 91 & 0.152 & 0 & 0.012 & \textbf{0.161} & 0 & 0 \\
reptilia-tortoise-network-bsv & 136 & 0.040 & 0 & 0 & 0.031 & \textbf{0.162} & 0.003 \\
\bottomrule
\end{tabular}
\end{table}

\section{Proof of main results}
Before proving Theorem \ref{thm1} and Theorem \ref{thm2}, we present two lemmas.

\begin{Lemma}\label{lem1}
Suppose $A\sim \mathcal{G}_{n}(m,r_n)$ with $r_n=o(1)$ and $nr_n^{m}=\omega(1)$. Then 
\begin{equation}\label{eedgfein}
    \mathbb{E}[A_{12}|X_1]=(2r_n)^{m}.
\end{equation}
\begin{equation}\label{etrinin}
    \mathbb{E}[A_{12}A_{23}A_{31}|X_1]=(3r_n^2)^m.
\end{equation}
\end{Lemma}

\noindent
{\bf Proof of Lemma \ref{lem1}:} Let $X_0=(\frac{1}{2}, \frac{1}{2},\dots,\frac{1}{2})\in[0,1]^m$ be a point with each coordinate component $\frac{1}{2}$.  Firstly, we prove (\ref{eedgfein}). Let $X_i=(X_{i1},X_{i2},\dots,X_{im})$. Then $X_{ik}$ ($1\leq i\leq n$, $1\leq k\leq m$) are independent and follow the uniform distribution $U[0,1]$.
 Then
\begin{eqnarray}\nonumber
    \mathbb{E}[A_{12}|X_1]&=&\mathbb{P}(d(X_1,X_2)\leq r_n|X_1=X_0)\\ \nonumber
    &=&\prod_{k=1}^m\mathbb{P}(\min\{|X_{0k}-X_{2k}|,1-|X_{0k}-X_{2k}|\}\leq r_n|X_1=X_0)\\ \label{lem1eq1}
    &=&\Big(\mathbb{P}(\min\{|X_{01}-X_{21}|,1-|X_{01}-X_{21}|\}\leq r_n|X_1=X_0)\Big)^m.
\end{eqnarray}
By assumption, $r_n=o(1)$, then
 $\min\{|X_{01}-X_{21}|,1-|X_{01}-X_{21}|\}\leq r_n$ is equivalent to 
\[X_{01}-r_n\leq X_{21}\leq X_{01}+r_n.\]
Then
\begin{equation*}
\mathbb{P}\left(\min\{|X_{01}-X_{21}|,1-|X_{01}-X_{21}|\}\leq r_n|X_{01}=\frac{1}{2}\right)=2r_n.
\end{equation*}
Hence, we have
\begin{eqnarray*} 
    \mathbb{E}[A_{12}|X_1]
    =(2r_n)^m.
\end{eqnarray*}

Next, we prove (\ref{etrinin}).
Since $X_1$ is uniformly distributed on the unit square $[0,1]^m$, then
\begin{eqnarray}\nonumber
\mathbb{E}[A_{12}A_{23}A_{31}|X_1]&=& \mathbb{E}[A_{12}A_{23}A_{31}|X_1=X_0]\\
&=&\mathbb{P}[A_{12}A_{23}A_{31}=1|X_1=X_0].
\end{eqnarray}
For $x=(x_1,x_2,\dots,x_m)$, define the ball around $x$ with radius $r_n$ as 
\[B_x(r_n)=\big\{(y_1,y_2,\dots,y_m)| x_j-r_n\leq y_j\leq x_j+r_n, j\in[m]\big\}.\]
Note that $A_{12}A_{23}A_{31}=1$ implies
$A_{12}=1$, $A_{23}=1$ and $A_{31}=1$.  Then $X_2,X_3\in B_{X_1}(r_n)=B_{X_0}(r_n)$ and $X_3\in B_{X_0}(r_n)\cap B_{X_2}(r_n)$, given $X_1=X_0$. $A_{12}=1$ if and only if $X_{2j}=\frac{1}{2}+\lambda_jr_n$ with $\lambda_j\in(-1,1)$ for all $j\in[m]$. 
Without loss of generality, let $\lambda_j\geq0$ for $1\leq j\leq k$ and $\lambda_j\leq0$ for $k+1\leq j\leq m$. Define a region $\Omega_2$ as
\[\Omega_2=\Big\{X_2\Big|X_{2j}=\frac{1}{2}+\lambda_jr_n,0\leq\lambda_1,\lambda_2, \dots,\lambda_k\leq1, -1\leq\lambda_{k+1},\lambda_{k+2},
\dots,\lambda_{m}\leq0\Big\}.\]
Given $X_2\in\Omega_2$, $X_3\in B_{X_0}(r_n)\cap B_{X_2}(r_n)$ is equivalent to
\begin{eqnarray*}
    X_{2j}-r_n&\leq& X_{3j}\leq\frac{1}{2}+r_n,\ \ \ \ j=1,2,\dots,k,\\
    \frac{1}{2}-r_n&\leq& X_{3j}\leq X_{2j}+r_n,\ \ \ \ j=k+1,k+2,\dots,m.
\end{eqnarray*}
Then
\begin{eqnarray*}
&&\mathbb{P}(X_2\in \Omega_2, X_3\in B_{X_0}(r_n)\cap B_{X_2}(r_n)|X_1=X_0)\\
&=&\int_{\Omega_2}dX_2\int_{B_{X_0}(r_n)\cap B_{X_2}(r_n)}dX_3\\
&=&\int_{\Omega_2}dX_2\left(\prod_{j=1}^k\big(\frac{1}{2}+2r_n-X_{2j}\big)\right)\left(\prod_{j=k+1}^m\big(2r_n+X_{2j}-\frac{1}{2}\big)\right)\\
&=&\left(\frac{3r_n^2}{2}\right)^k\left(\frac{3r_n^2}{2}\right)^{m-k}\\
&=&\left(\frac{3r_n^2}{2}\right)^m.
\end{eqnarray*}

There are $2^m$ such subsets as $\Omega_2$. Then
\begin{eqnarray*}
&&\mathbb{P}(X_2\in B_{X_0}(r_n), X_3\in B_{X_0}(r_n)\cap B_{X_2}(r_n)|X_1=X_0)=2^m\left(\frac{3r_n^2}{2}\right)^m=(3r_n^2)^m.
\end{eqnarray*}
The proof is complete.

\qed

Now we present an asymptotic result of degenerate U-statistics.
Let $\psi_n=\psi_n(x_1,x_2,\dots,x_k)$ be a symmetric function that may depend on $n$.  Let $X_1,X_2,\dots,X_n$ be an i.i.d. sample from some distribution $F$. The U-statistic of order $k$ with kernel $\psi_n$ is defined as
\[U_n=\frac{1}{\binom{n}{k}}\sum_{1\leq i_1<i_2<\dots<i_k\leq n}\psi_n(X_{i_1},X_{i_2},\dots,X_{i_k}).\]
Assume $\mathbb{E}[\psi_n(X_1,X_2,\dots,X_k)]=0$. Define
\[\psi_{n,l}(X_1,\dots,X_l)=\mathbb{E}[\psi_n(X_{1},X_{2},\dots,X_{k})|X_1,\dots,X_l],\ \ \ \ 1\leq l\leq k,\]
\[\sigma_{nl}^2=Var(\psi_{n,l}(X_1,\dots,X_l)), \ \ \ \ 1\leq l\leq k,\]
and
\[G(x,y)=\mathbb{E}[\psi_{n,2}(x,X_1)\psi_{n,2}(y,X_1)].\]
If $\psi_{n,1}=0$, the U-statistic $U_n$ is  degenerate. The following lemma   provides the asymptotic distribution of degenerate U-statistic $U_n$ (Lemma B.4 in \cite{FL96}).

\begin{Lemma}\label{lem4}
    Suppose $\psi_{n,1}=0$ and $\sigma_{nk}^2<\infty$ for each $n$. If $\sigma_{nl}^2=o(n^{l-2}\sigma_{n2}^2)$ for $l\geq 3$ and 
    \begin{equation}\label{lem4eq1}
        \frac{\mathbb{E}[G(X_1,X_2)^2]+n^{-1}\mathbb{E}[\psi_{n,2}(X_1,X_2)^4]}{\left(\mathbb{E}[\psi_{n,2}(X_1,X_2)^2]\right)^2}=o(1),
    \end{equation}
    then
      \begin{equation*}
\frac{\sqrt{2}nU_n}{k(k-1)\sigma_{n2}}\Rightarrow N(0,1).
        \end{equation*}
\end{Lemma}

\subsection{Proof of Proposition \ref{prop}}

By Lemma \ref{lem1}, we have
\begin{eqnarray}\nonumber
\mathbb{E}\left[A_{12}A_{13}\right]&=&\mathbb{E}\big[\mathbb{E}\left[A_{12}A_{13}|X_1\right]\big]\\ \nonumber
&=&\mathbb{E}\big[\mathbb{E}\left[A_{12}|X_1\right]\mathbb{E}\left[A_{13}|X_1\right]\big]\\ \label{twopathe}
&=&(2r_n)^{2m}.
\end{eqnarray}
Then
\begin{eqnarray}\label{proppreq1}
\sum_{i\neq j\neq k}\mathbb{E}\left[A_{ij}A_{jk}\right]&=&n(n-1)(n-2)(2r_n)^{2m},\\ \label{proppreq2}
\sum_{i\neq j\neq k}\mathbb{E}\left[A_{ij}A_{jk}A_{ki}\right]&=&n(n-1)(n-2)(3r_n^2)^m.
\end{eqnarray}

Next, we prove the first term and second term of $D_n$ in (\ref{dn}) converges to their expectations in (\ref{proppreq1}) and (\ref{proppreq2}), respectively. To this end, we will show their variances are of orders smaller than their expectations.

Firstly, we consider the second term  of $D_n$ in (\ref{dn}). Its variance is equal to
\begin{eqnarray}\nonumber
&&\mathbb{E}\left[\left(\sum_{i\neq j\neq k}\big(A_{ij}A_{ik}-\mathbb{E}[A_{ij}A_{ik}]\big)\right)^2\right]\\ \label{secondpathm}
&=&\sum_{\substack{i\neq j\neq k\\  i_1\neq j_1\neq k_1}}\mathbb{E}\left[\big(A_{ij}A_{ik}-\mathbb{E}[A_{ij}A_{ik}]\big)\big(A_{i_1j_1}A_{i_1k_1}-\mathbb{E}[A_{i_1j_1}A_{i_1k_1}]\big)\right].
\end{eqnarray}
If $\{i,j,k\}\cap\{i_1,j_1,k_1\}=\emptyset$, then $A_{ij}A_{ki}$ and $A_{i_1j_1}A_{k_1i_1}$ are independent. In this case,
 \begin{equation}\label{propeqp}
 \mathbb{E}\left[\big(A_{ij}A_{ik}-\mathbb{E}[A_{ij}A_{ik}]\big)\big(A_{i_1j_1}A_{i_1k_1}-\mathbb{E}[A_{i_1j_1}A_{i_1k_1}]\big)\right]=0.
 \end{equation}
Suppose $|\{i,j,k,i_1,j_1,k_1\}|=5$.
Without loss of generality, let $i=i_1$. By Lemma \ref{lem1}, we have
 \begin{eqnarray}\nonumber
\mathbb{E}\left[A_{ij}A_{ik}A_{i_1j_1}A_{i_1k_1}\right]
&=&\mathbb{E}\left[A_{ij}A_{ik}A_{ij_1}A_{ik_1}\right]\\ \nonumber
&=&\mathbb{E}\big[\mathbb{E}\left[A_{ij}A_{ik}A_{ij_1}A_{ik_1}|X_i\right]\big]\\ \nonumber
&=&\mathbb{E}\big[\mathbb{E}\left[A_{ij}|X_i\right]\mathbb{E}\left[A_{ik}|X_i\right]\mathbb{E}\left[A_{ij_1}|X_i\right]\mathbb{E}\left[A_{ik_1}|X_i\right]\big]\\ \label{lempropeqp}
&=&(2r_n)^{4m}.
 \end{eqnarray}
By (\ref{twopathe}) and (\ref{lempropeqp}), (\ref{propeqp}) holds.

Suppose $|\{i,j,k,i_1,j_1,k_1\}|=4$.  There are three scenarios: (a) $i_1,j_1\in \{i,j,k\}$, (b)  $i_1,k_1\in \{i,j,k\}$ and (c) $j_1,k_1\in \{i,j,k\}$. If (a) $i_1,j_1\in \{i,j,k\}$, then
 \begin{eqnarray}\nonumber
 &&\mathbb{E}\left[\big(A_{ij}A_{ik}-\mathbb{E}[A_{ij}A_{ik}]\big)\big(A_{i_1j_1}A_{i_1k_1}-\mathbb{E}[A_{i_1j_1}A_{i_1k_1}]\big)\right]\\ \nonumber
 &\leq&\mathbb{E}\left[A_{ij}A_{ik} A_{i_1k_1}\right]\\ \label{propeqp1}
 &=&O(r_n^{3m}).
 \end{eqnarray}
 Here, the last equality is obtained by a similar argument as in (\ref{twopathe}) and (\ref{lempropeqp}).
Similarly, (\ref{propeqp1}) holds for case (b) and (c).

Suppose $|\{i,j,k,i_1,j_1,k_1\}|=3$. In this case,  $i_1,j_1,k_1\in\{i,j,k\}$. Then
 \begin{eqnarray}\nonumber
 &&\mathbb{E}\left[\big(A_{ij}A_{ik}-\mathbb{E}[A_{ij}A_{ik}]\big)\big(A_{i_1j_1}A_{i_1k_1}-\mathbb{E}[A_{i_1j_1}A_{i_1k_1}]\big)\right]\\ \nonumber
 &\leq&\mathbb{E}\left[A_{ij}A_{ik} \right]\\ \label{propeqp2}
 &=&O(r_n^{2m}).
 \end{eqnarray}

By (\ref{secondpathm})- (\ref{propeqp2}), we have
\begin{eqnarray*}
\mathbb{E}\left[\left(\sum_{i\neq j\neq k}\big(A_{ij}A_{ik}-\mathbb{E}[A_{ij}A_{ik}]\big)\right)^2\right]=O(n^4r_n^{3m}+n^3r_n^{2m}),
\end{eqnarray*}
which implies
\begin{eqnarray}\label{propppeq1}
\sum_{i\neq j\neq k}A_{ij}A_{ik}=\sum_{i\neq j\neq k}\mathbb{E}[A_{ij}A_{ik}]=n(n-1)(n-2)(2r_n)^{2m}\left(1+O_P\left(\frac{1}{n\sqrt{r_n^m}}\right)\right).
\end{eqnarray}

Now, we consider the first term  of $D_n$ in (\ref{dn}). Its variance is equal to
\begin{eqnarray}\nonumber
    &&\mathbb{E}\left[\left(\sum_{i\neq j\neq k}\big(A_{ij}A_{jk}A_{ki}-\mathbb{E}[A_{ij}A_{jk}A_{ki}]\big)\right)^2\right]\\ \label{secondtrimo}
    &=&\sum_{\substack{i\neq j\neq k\\ i_1\neq j_1\neq k_1}}\mathbb{E}\left[\big(A_{ij}A_{jk}A_{ki}-\mathbb{E}[A_{ij}A_{jk}A_{ki}]\big)\big(A_{i_1j_1}A_{j_1k_1}A_{k_1i_1}-\mathbb{E}[A_{i_1j_1}A_{j_1k_1}A_{k_1i_1}]\big)\right].
\end{eqnarray}
If $\{i,j,k\}\cap\{i_1,j_1,k_1\}=\emptyset$, then $A_{ij}A_{jk}A_{ki}$ and $A_{i_1j_1}A_{j_1k_1}A_{k_1i_1}$ are independent. In this case,
\begin{equation}\label{propeq1}
\mathbb{E}\left[\big(A_{ij}A_{jk}A_{ki}-\mathbb{E}[A_{ij}A_{jk}A_{ki}]\big)\big(A_{i_1j_1}A_{j_1k_1}A_{k_1i_1}-\mathbb{E}[A_{i_1j_1}A_{j_1k_1}A_{k_1i_1}]\big)\right]=0.
\end{equation}
Suppose $|\{i,j,k,i_1,j_1,k_1\}|=5$. By a similar argument as in (\ref{twopathe}) and (\ref{lempropeqp}), (\ref{propeq1}) holds.

Suppose $|\{i,j,k,i_1,j_1,k_1\}|=4$. Without loss of generality, let $i_1,j_1\in\{i,j,k\}$. Then
\begin{eqnarray}\nonumber
&&\mathbb{E}\left[\big(A_{ij}A_{jk}A_{ki}-\mathbb{E}[A_{ij}A_{jk}A_{ki}]\big)\big(A_{i_1j_1}A_{j_1k_1}A_{k_1i_1}-\mathbb{E}[A_{i_1j_1}A_{j_1k_1}A_{k_1i_1}]\big)\right]\\ \nonumber
&\leq&\mathbb{E}\left[A_{ij}A_{jk}A_{ki}A_{i_1j_1}A_{j_1k_1}A_{k_1i_1}\right]\\ \nonumber
&=&\mathbb{E}\left[A_{ij}A_{jk}A_{ki}A_{jk_1}A_{k_1i}\right]\\ \nonumber
&\leq&\mathbb{E}\left[A_{jk}A_{ki}A_{jk_1}\right]\\ \label{propeq2}
&=&O(r_n^{3m}).
\end{eqnarray}
Here, the last equality is obtained  by a similar argument as in (\ref{lempropeqp}).

Suppose $|\{i,j,k,i_1,j_1,k_1\}|=3$. Then $\{i_1,j_1,k_1\}=\{i,j,k\}$. In this case,
\begin{eqnarray}\nonumber
&&\mathbb{E}\left[\big(A_{ij}A_{jk}A_{ki}-\mathbb{E}[A_{ij}A_{jk}A_{ki}]\big)\big(A_{i_1j_1}A_{j_1k_1}A_{k_1i_1}-\mathbb{E}[A_{i_1j_1}A_{j_1k_1}A_{k_1i_1}]\big)\right]\\ \nonumber
&\leq&\mathbb{E}\left[A_{ij}A_{jk}A_{ki}A_{i_1j_1}A_{j_1k_1}A_{k_1i_1}\right]\\ \nonumber
&=&\mathbb{E}\left[A_{ij}A_{jk}A_{ki}\right]\\ \nonumber
&\leq&\mathbb{E}\left[A_{jk}A_{ki}\right]\\ \label{Mpropeq2}
&=&O(r_n^{2m}).
\end{eqnarray}

Combining (\ref{secondtrimo})-(\ref{Mpropeq2}) yields
\begin{eqnarray*}
  \mathbb{E}\left[\left(\sum_{i\neq j\neq k}\big(A_{ij}A_{jk}A_{ki}-\mathbb{E}[A_{ij}A_{jk}A_{ki}]\big)\right)^2\right]=O(n^4r_n^{3m}+n^3r_n^{2m}).
\end{eqnarray*}
Hence,
\begin{equation}\label{propeqt}
\sum_{i\neq j\neq k}A_{ij}A_{jk}A_{ki}=\sum_{i\neq j\neq k}\mathbb{E}[A_{ij}A_{jk}A_{ki}]=n(n-1)(n-2)(3r_n^2)^m\left(1+O_P\left(\frac{1}{n\sqrt{r_n^m}}\right)\right).
\end{equation}

By (\ref{propppeq1}) and (\ref{propeqt}), we have
\begin{eqnarray*}\nonumber
D_n&=&n(n-1)(n-2)(3r_n^2)^m\left[1+O_P\left(\frac{1}{n\sqrt{r_n^m}}\right)\right]\\
&&-n(n-1)(n-2)\left(\frac{3}{4}\right)^{m_0}(2r_n)^{2m}\left[1+O_P\left(\frac{1}{n\sqrt{r_n^m}}\right)\right].
\end{eqnarray*}
Under the null hypothesis $H_0$, $m=m_0$. Then
\[\left(\frac{3}{4}\right)^{m_0}(2r_n)^{2m_0}=(3r_n^2)^{m_0},\]
which implies
 \begin{eqnarray}\nonumber
D_n=O_P\left(\frac{n(n-1)(n-2)(3r_n^2)^m}{n\sqrt{r_n^m}}\right)=O_P\left(\frac{n^3r_n^{2m}}{n\sqrt{r_n^m}}\right).
\end{eqnarray}

Under the alternative hypothesis $H_1$, $m\neq m_0$. Then
\[\left(\frac{3}{4}\right)^{m_0}(2r_n)^{2m}\neq(3r_n^2)^{m},\]
which implies
\begin{eqnarray*}\nonumber
D_n=n(n-1)(n-2)r_n^{2m}\left[3^m-\left(\frac{3}{4}\right)^{m_0}4^{m}\right]+O_P\left(\frac{n^3r_n^{2m}}{n\sqrt{r_n^m}}\right).
\end{eqnarray*}

Then the proof is complete.

\qed

\subsection{Proof of Theorem \ref{thm1}}
Assume the null hypothesis $H_0$ holds; that is, $m=m_0$.
We will express $D_n$ as a U-statistic and apply the asymptotic theory of U-statistics to derive its asymptotic distribution.
To this end, define a function $h=h(X_1,X_2,X_3)$ as follows:
\begin{eqnarray} \nonumber
h(X_1,X_2,X_3)&=&6A_{12}A_{23}A_{31}-2\left(\frac{3}{4}\right)^{m_0}\big(A_{12}A_{23}+A_{12}A_{13}+A_{23}A_{31}\big)\\ \nonumber
&=&6I[d(X_1,X_2)<r_n]I[d(X_2,X_3)<r_n]I[d(X_3,X_1)<r_n]\\  \nonumber
&&-2\left(\frac{3}{4}\right)^{m_0}\Big(I[d(X_1,X_2)<r_n]I[d(X_2,X_3)<r_n]\\  \nonumber
&&+I[d(X_1,X_2)<r_n]I[d(X_3,X_1)<r_n]\\ \label{hfuc}
&&+I[d(X_2,X_3)<r_n]I[d(X_3,X_1)<r_n]\Big) .
\end{eqnarray}
 Then $h(X_1,X_2,X_3)$ is symmetric and
\begin{eqnarray}
D_n=\binom{n}{3}U_n,
\end{eqnarray}
where $U_n$ is a U-statistic defined as
\[U_n=\frac{1}{\binom{n}{3}
}\sum_{i<j<k}h(X_i,X_j,X_k).\]

Next we show $U_n$ is a degenerate U-statistic and derive its asymptotic distribution. By Lemma \ref{lem1},
it is easy to verify that
\[\mathbb{E}[h(X_1,X_2,X_3)]=0,\ \ \ \ \ \ \ \mathbb{E}[h(X_1,X_2,X_3)|X_1]=0.\]

Let
\begin{equation}
g(X_1,X_2)=\mathbb{E}[h(X_1,X_2,X_3)|X_1,X_2].
\end{equation}
Clearly, $\mathbb{E}[g(X_1,X_2)]=0$.
Next, we show the variance of $g(X_1,X_2)$ is not equal to zero.
Let $\sigma_{n2}^2=\mathbb{E}[g(X_1,X_2)^2]$.

By (\ref{hfuc}), it is easy to verify that
\begin{eqnarray} \label{hger}
h(X_1,X_2,X_3)I[d(X_1,X_2)\geq r_n]=-2\left(\frac{3}{4}\right)^{m_0}A_{23}A_{31}I[d(X_1,X_2)\geq r_n],
\end{eqnarray}
and
\begin{eqnarray} \nonumber
&&h(X_1,X_2,X_3)I[d(X_1,X_2)< r_n]\\  \label{hler}
&=&\Big[6-2\left(\frac{3}{4}\right)^{m_0}\Big]A_{12}A_{23}A_{31}-2\left(\frac{3}{4}\right)^{m_0}\Big(A_{12}A_{23}+A_{12}A_{31}\Big).
\end{eqnarray}
Then the function $h(X_1,X_2,X_3)$ can be written as
\begin{eqnarray}\nonumber
&&h(X_1,X_2,X_3)\\ \label{hgeqleq}
&=&h(X_1,X_2,X_3)I[d(X_1,X_2)\geq r_n]+h(X_1,X_2,X_3)I[d(X_1,X_2)< r_n].
\end{eqnarray}
Therefore, $g(X_1,X_2)$ can be expressed as
\begin{eqnarray} \nonumber
g(X_1,X_2)&=&\mathbb{E}[h(X_1,X_2,X_3)|X_1,X_2]\\   \nonumber
&=&\mathbb{E}[h(X_1,X_2,X_3)|X_1,X_2]I[d(X_1,X_2)\geq r_n]\\ \label{hgeqleq2}
&&+\mathbb{E}[h(X_1,X_2,X_3)|X_1,X_2]I[d(X_1,X_2)< r_n].
\end{eqnarray}
Then the variance of $g(X_1,X_2)$ is equal to
\begin{eqnarray}\nonumber
\sigma_{n2}^2
&=&\mathbb{E}[g(X_1,X_2)^2]\\   \nonumber
&=&\mathbb{E}\Big[\Big(\mathbb{E}[h(X_1,X_2,X_3)|X_1,X_2]I[d(X_1,X_2)\geq r_n]\Big)^2\Big]\\ \label{0varexpress}
&&+\mathbb{E}\Big[\Big(\mathbb{E}[h(X_1,X_2,X_3)|X_1,X_2]I[d(X_1,X_2)< r_n]\Big)^2\Big].
\end{eqnarray}
By (\ref{hger}) and the propoerties of conditional expectation, it is easy to verify that
\begin{eqnarray}\nonumber
&&\mathbb{E}\Big[\Big(\mathbb{E}[h(X_1,X_2,X_3)|X_1,X_2]I[d(X_1,X_2)\geq r_n]\Big)^2\Big]\\ \nonumber
&=&\mathbb{E}\Big[\Big(\mathbb{E}[h(X_1,X_2,X_3)I[d(X_1,X_2)\geq r_n]|X_1,X_2]\Big)^2\Big]\\ \nonumber
&=&4\left(\frac{3}{4}\right)^{2m_0}\mathbb{E}\Big[\Big(\mathbb{E}\big[A_{23}A_{31}I[d(X_1,X_2)\geq r_n]|X_1,X_2\big]\Big)^2\Big]\\ \nonumber
&=&4\left(\frac{3}{4}\right)^{2m_0}\mathbb{E}\Big[\mathbb{E}\big[A_{23}A_{31}I[d(X_1,X_2)\geq r_n]|X_1,X_2\big]\mathbb{E}\big[A_{24}A_{41}I[d(X_1,X_2)\geq r_n]|X_1,X_2\big]\Big]\\ \nonumber
&=&4\left(\frac{3}{4}\right)^{2m_0}\mathbb{E}\Big[\mathbb{E}\Big[A_{23}A_{31}A_{24}A_{41}I[d(X_1,X_2)\geq r_n]|X_1,X_2\Big]\Big]\\ \label{varexpress}
&=&4\left(\frac{3}{4}\right)^{2m_0}\mathbb{E}\Big[A_{23}A_{31}A_{24}A_{41}I[d(X_1,X_2)\geq r_n]\Big].
\end{eqnarray}
Note that 
\[I[d(X_1,X_2)\geq r_n]\geq I[r_n\leq d(X_1,X_2)\leq 1.2r_n].\]
If $X_3,X_4\in B_{\frac{X_1+X_2}{2}}\big(\frac{r_n}{4}\big)$, then
\[ d(X_1,X_3)\leq d\left(X_1,\frac{X_1+X_2}{2}\right)+d\left(\frac{X_1+X_2}{2},X_3\right)<0.6r_n+\frac{r_n}{4}<r_n.\]
Similarly,
\[d(X_2,X_3)<r_n,\ \ \ d(X_1,X_4)<r_n,\ \ \ \ d(X_2,X_4)<r_n.\]
Then
\begin{eqnarray}\nonumber
   && \mathbb{E}\Big[A_{23}A_{31}A_{24}A_{41}I[d(X_1,X_2)\geq r_n]\Big]\\ \nonumber
   &\geq&\mathbb{E}\Big[I[d\left(X_3,\frac{X_1+X_2}{2}\right)<\frac{r_n}{4}]I[d\left(X_3,\frac{X_1+X_2}{2}\right)<\frac{r_n}{4}]I[r_n\leq d(X_1,X_2)\leq 1.2r_n]\Big]\\ \nonumber
   &=&\left(\frac{r_n}{2}\right)^{2m_0}\big[(2.4r_n)^{m_0}-(2r_n)^{m_0})\big]\\ \label{lagelow}
   &=&\frac{2.4^{m_0}-2^{m_0}}{4^{m_0}}r_n^{3m_0}.
\end{eqnarray}

Combining  (\ref{0varexpress}), (\ref{varexpress}) and (\ref{lagelow}) yields
\begin{eqnarray}\label{varlowerb}
\sigma_{n2}^2\geq \frac{3^{2m_0}\big(2.4^{m_0}-2^{m_0}\big)}{4^{3m_0-1}} r_n^{3m_0}.
\end{eqnarray}

Next, we get an upper bound of $\sigma_{n2}^2$. Note that
\begin{eqnarray}\label{e36q}
\mathbb{E}\Big[A_{23}A_{31}A_{24}A_{41}I[d(X_1,X_2)\geq r_n]\Big]
\leq\mathbb{E}\Big[A_{23}A_{24}A_{41}\Big]=(2r_n)^{3m_0}.
\end{eqnarray}
Moreover, by (\ref{hler}), we have
\begin{eqnarray} \nonumber
&&\mathbb{E}[h(X_1,X_2,X_3)I[d(X_1,X_2)< r_n]|X_1,X_2]\\  \label{0hler}
&=&\Big[6-2\left(\frac{3}{4}\right)^{m_0}\Big]\mathbb{E}[A_{12}A_{23}A_{31}|X_1,X_2]-4(2r_n)^{m_0}\left(\frac{3}{4}\right)^{m_0}A_{12}.
\end{eqnarray}
Then
\begin{eqnarray} \nonumber
&&\mathbb{E}\Big[\Big(\mathbb{E}\big[h(X_1,X_2,X_3)I[d(X_1,X_2)< r_n]|X_1,X_2\big]\Big)^2\Big]\\ \nonumber
&\leq&2\Big[6-2\left(\frac{3}{4}\right)^{m_0}\Big]^2\mathbb{E}\Big[\big(\mathbb{E}[A_{12}A_{23}A_{31}|X_1,X_2]\big)^2\Big]+32(2r_n)^{2m_0}\left(\frac{3}{4}\right)^{2m_0}\mathbb{E}[A_{12}]\\ \nonumber
&=&2\Big[6-2\left(\frac{3}{4}\right)^{m_0}\Big]^2\mathbb{E}\Big[A_{12}A_{23}A_{31}A_{24}A_{41}\Big]+32\left(\frac{3}{4}\right)^{2m_0}(2r_n)^{3m_0}\\ \nonumber
&\leq&2\Big[6-2\left(\frac{3}{4}\right)^{m_0}\Big]^2\mathbb{E}\Big[A_{23}A_{31}A_{24}\Big]+32\left(\frac{3}{4}\right)^{2m_0}(2r_n)^{3m_0}\\ \label{varhupper}
&=&2\Big[6-2\left(\frac{3}{4}\right)^{m_0}\Big]^2(2r_n)^{3m_0}+32\left(\frac{3}{4}\right)^{2m_0}(2r_n)^{3m_0}.
\end{eqnarray}
Combining (\ref{0varexpress}), (\ref{varexpress}), (\ref{e36q}), (\ref{0hler}) and (\ref{varhupper}) yields
\begin{equation}\label{var3m0}
\sigma_{n2}^2=O(r_n^{3m_0}).
\end{equation}

By (\ref{varlowerb}) and (\ref{var3m0}), we get
\begin{equation}\label{varorderder}
\sigma_{n2}^2=\Theta\left( r_n^{3m_0}\right).
\end{equation}
Hence, $U_n$ is a degenerated U-statistic with $\sigma_{n2}^2>0$.

Next, we will use Lemma \ref{lem4} to derive the asymptotic distribution of the degenerated U-statistic $U_n$.

Let $\sigma_{n3}^2=Var(h(X_1,X_2,X_3))$. Firstly, we verify the condition $\sigma_{n3}^2=o(n\sigma_{n2}^2)$. Note that
\[(a+b+c+d)^2\leq 4(a^2+b^2+c^2+d^2),\]
where $a,b,c,d$ are real numbers.
Then, by (\ref{hfuc}), it is easy to verify that
\begin{eqnarray}\label{sigam3n}
\sigma_{n3}^2\leq 144\mathbb{E}[A_{12}A_{13}A_{23}]+16\left(\frac{3}{4}\right)^{2m_0}\mathbb{E}[A_{12}A_{13}]=O\left( r_n^{2m_0}\right).
\end{eqnarray}
Combining (\ref{varorderder}), (\ref{sigam3n}) and the assumption that $nr_n^{m_0}=\omega(1)$ yields
\begin{eqnarray}\label{lemcon1}
\frac{\sigma_{n3}^2}{n\sigma_{n2}^2}=O\left(\frac{r_n^{2m_0}}{nr_n^{3m_0}}\right)=O\left(\frac{1}{nr_n^{m_0}}\right)=o(1).
\end{eqnarray}

Next, we verify condition (\ref{lem4eq1}). By (\ref{hger}), one has
\begin{eqnarray} \Big(h(X_1,X_2,X_3)I[d(X_1,X_2)\geq r_n]\Big)^4
\leq2^4\left(\frac{3}{4}\right)^{4m_0}A_{23}A_{31}.
\end{eqnarray}
Then
\begin{eqnarray} \nonumber
&&\mathbb{E}\Bigg[\Big(\mathbb{E}\big[h(X_1,X_2,X_3)I[d(X_1,X_2)\geq r_n]|X_1,X_2\big]\Big)^4\Bigg]\\ \nonumber
&\leq&2^4\left(\frac{3}{4}\right)^{4m_0}\mathbb{E}\Bigg[\Big(\mathbb{E}\big[A_{23}A_{31}|X_1,X_2\big]\Big)^4\Bigg]\\ \nonumber
&=&2^4\left(\frac{3}{4}\right)^{4m_0}\mathbb{E}\Big[A_{23}A_{31}A_{24}A_{41}A_{25}A_{51}A_{26}A_{61}\Big]\\ \nonumber
&\leq&2^4\left(\frac{3}{4}\right)^{4m_0}\mathbb{E}\Big[A_{23}A_{31}A_{41}A_{51}A_{61}\Big]\\ \label{moneq1}
&=&O(r_n^{5m_0}).
\end{eqnarray}
Note that
\[(a+b)^4\leq 16(a^4+b^4),\]
where $a, b$ are real numbers.
By (\ref{0hler}), we have
\begin{eqnarray} \nonumber
&&\mathbb{E}\Big[\Big(\mathbb{E}\big[h(X_1,X_2,X_3)I[d(X_1,X_2)< r_n]|X_1,X_2\big]\Big)^4\Big]\\ \nonumber
&\leq&16\Big[6-2\left(\frac{3}{4}\right)^{m_0}\Big]^4\mathbb{E}\Big[\Big(\mathbb{E}\big[A_{23}A_{31}|X_1,X_2\big]\Big)^4\Big]+4^6\left(\frac{3}{4}\right)^{4m_0}(2r_n)^{4m_0}\mathbb{E}[A_{12}]\\ \label{moneq2}
&=&O(r_n^{5m_0}).
\end{eqnarray}

By (\ref{hgeqleq2}), (\ref{moneq1}) and (\ref{moneq2}), we get
\[\mathbb{E}[g(X_1,X_2)^4]=O(r_n^{5m_0}).\]
By (\ref{varorderder}),
$\sigma_{n2}^2=\Theta\left( r_n^{3m_0}\right)$. By assumption, $nr_n^{m_0}=\omega(1)$. 
Hence,
\begin{equation}\label{eg4mo}
\frac{\mathbb{E}[g(X_1,X_2)^4]}{n\sigma_{2n}^4}=O\left(\frac{r_n^{5m_0}}{nr_n^{6m_0}}\right)=O\left(\frac{1}{nr_n^{m_0}}\right)=o(1).
\end{equation}

Now we get an upper bound of $\mathbb{E}[G(X_3,X_4)^2]$.
By definition, we have
\begin{eqnarray}\nonumber
    &&\mathbb{E}[G(X_3,X_4)^2]\\ \nonumber
    &=&\mathbb{E}\Big[\Big(\mathbb{E}[g(X_3,X_1)g(X_1,X_4)|X_3,X_4]\Big)^2 \Big]\\ \nonumber
    &=&\mathbb{E}\Big[\mathbb{E}[g(X_3,X_1)g(X_1,X_4)|X_3,X_4]\mathbb{E}[g(X_3,X_2)g(X_2,X_4)|X_3,X_4] \Big]\\ \nonumber
    &=&\mathbb{E}\Big[\mathbb{E}[g(X_3,X_1)g(X_1,X_4)g(X_3,X_2)g(X_2,X_4)|X_3,X_4] \Big]\\ \label{egxys}
    &=&\mathbb{E}[g(X_3,X_1)g(X_4,X_1)g(X_3,X_2)g(X_4,X_2)].
\end{eqnarray}

By (\ref{0hler}), we have
\begin{eqnarray} \nonumber
&&\big|\mathbb{E}[h(X_1,X_2,X_3)I[d(X_1,X_2)< r_n]|X_1,X_2]\big|\\    \nonumber
&=&\Big|\Big[6-2\left(\frac{3}{4}\right)^{m_0}\Big]\mathbb{E}[A_{12}A_{23}A_{31}|X_1,X_2]-4(2r_n)^{m_0}\left(\frac{3}{4}\right)^{m_0}A_{12}\Big|\\ \label{00hler}
&\leq&C_1\mathbb{E}[A_{23}A_{31}|X_1,X_2]+C_2A_{12},
\end{eqnarray}
where
\[C_1=6-2\left(\frac{3}{4}\right)^{m_0}>0,\ \ \ \ \ C_2=4\left(\frac{3}{4}\right)^{m_0}2^{m_0}.\]

By (\ref{hger}), we have
\begin{eqnarray} \nonumber
&&\big |\mathbb{E}\big[h(X_1,X_2,X_3)I[d(X_1,X_2)\geq r_n]|X_1,X_2\big]\geq r_n]\big |\\ \nonumber
&\leq& 2\left(\frac{3}{4}\right)^{m_0}\mathbb{E}[A_{23}A_{31}|X_1,X_2]I[d(X_1,X_2)\geq r_n]\\  \label{0hger1}
&\leq&C_3\mathbb{E}[A_{23}A_{31}|X_1,X_2],
\end{eqnarray}
where
\[C_3=2\left(\frac{3}{4}\right)^{m_0}.\]

By (\ref{hgeqleq2}), (\ref{00hler}) and (\ref{0hger1}), we get
\begin{eqnarray} \nonumber
\big | g(X_1,X_2)\big |&=&\big |\mathbb{E}[h(X_1,X_2,X_3)|X_1,X_2]\big |\\ \nonumber
&=&\big |\mathbb{E}[h(X_1,X_2,X_3)|X_1,X_2]I[d(X_1,X_2)\geq r_n]\\ \nonumber
&&+\mathbb{E}[h(X_1,X_2,X_3)|X_1,X_2]I[d(X_1,X_2)< r_n]\big |\\ \nonumber
&\leq&\big |\mathbb{E}[h(X_1,X_2,X_3)|X_1,X_2]I[d(X_1,X_2)\geq r_n]\big |\\ \nonumber
&&+\big |\mathbb{E}[h(X_1,X_2,X_3)|X_1,X_2]I[d(X_1,X_2)< r_n]\big |\\ \label{absgxy}
&\leq&6\mathbb{E}\big[A_{23}A_{31}|X_1,X_2\big]+C_2r_n^{m_0}A_{12}.
\end{eqnarray}

Combining (\ref{egxys}) and (\ref{absgxy}) yields
\begin{eqnarray}\nonumber
    &&\mathbb{E}[G(X_3,X_4)^2]\\ \nonumber
    &\leq&\mathbb{E}[|g(X_3,X_1)||g(X_4,X_1)||g(X_3,X_2)||g(X_4,X_2)|]\\ \label{egsquper}
&\leq&\mathbb{E}\Big[\big(6F_{13}+C_2r_n^{m_0}A_{13}\big) \big(6F_{14}+C_2r_n^{m_0}A_{14}\big) \big(6F_{23}+C_2r_n^{m_0}A_{23}\big) \big(6F_{24}+C_2r_n^{m_0}A_{24}\big)\Big],
\end{eqnarray}
where
\[F_{12}=\mathbb{E}\big[A_{23}A_{31}|X_1,X_2\big].\]

Note that
\begin{eqnarray}\label{egsquper1}
\mathbb{E}\big[C_2^4r_n^{4m_0}A_{13}A_{14}A_{23}A_{24}\big]
    \leq\mathbb{E}\big[C_2^4r_n^{4m_0}A_{13}A_{14}A_{23}\big]=O(r_n^{7m_0}),
\end{eqnarray}

\begin{eqnarray}\nonumber
\mathbb{E}\Big[6^4F_{13}F_{14}F_{23}F_{24}\Big]   &=&\mathbb{E}\Bigg[6^4\mathbb{E}\Big[A_{15}A_{53}A_{16}A_{64}A_{37}A_{72}A_{48}A_{82}\Big|X_1,X_2,X_3,X_4\Big]\Bigg]\\ \nonumber
&=&6^4\mathbb{E}\Big[A_{15}A_{53}A_{16}A_{64}A_{37}A_{72}A_{48}A_{82}\Big]\\ \nonumber
&\leq&6^4\mathbb{E}\Big[A_{15}A_{53}A_{16}A_{64}A_{37}A_{72}A_{48}\Big]\\ \label{egsquper2}
    &=&O(r_n^{7m_0}),
\end{eqnarray}

\begin{eqnarray}\nonumber
&&\mathbb{E}\Big[6^3C_2r_n^{m_0}A_{13} F_{14}F_{23}F_{24}\Big]\\ \nonumber
    &=&6^3C_2r_n^{m_0}\mathbb{E}\Bigg[\mathbb{E}\Big[A_{13}A_{15}A_{54}A_{26}A_{63}A_{27}A_{74}\Big|X_1,X_2,X_3,X_4\Big]\Bigg]\\ \nonumber
&=&6^3C_2r_n^{m_0}\mathbb{E}\Big[A_{13}A_{15}A_{54}A_{26}A_{63}A_{27}A_{74}\Big]\\ \nonumber
&\leq&6^3C_2r_n^{m_0}\mathbb{E}\Big[A_{13}A_{15}A_{54}A_{26}A_{63}A_{27}\Big]\\ \label{egsquper3}
    &=&O(r_n^{7m_0}),
\end{eqnarray}

Similarly, we have
\begin{eqnarray}\label{egsquper4}
\mathbb{E}\Big[6^3C_2r_n^{m_0}F_{13} A_{14}F_{23}F_{24}\Big]&=&O(r_n^{7m_0}),\\ \label{egsquper5}
\mathbb{E}\Big[6^3C_2r_n^{m_0}F_{13} F_{14}A_{23}F_{24}\Big]&=&O(r_n^{7m_0}),\\ \label{egsquper6}
\mathbb{E}\Big[6^3C_2r_n^{m_0}F_{13} F_{14}F_{23}A_{24}\Big]&=&O(r_n^{7m_0}).
\end{eqnarray}

\begin{eqnarray}\nonumber
&&\mathbb{E}\Big[6^2C_2^2r_n^{2m_0}A_{13} A_{14}F_{23}F_{24}\Big]\\ \nonumber
    &=&6^2C_2^2r_n^{2m_0}\mathbb{E}\Bigg[\mathbb{E}\Big[A_{13}A_{14}A_{26}A_{63}A_{27}A_{74}\Big|X_1,X_2,X_3,X_4\Big]\Bigg]\\ \nonumber
&=&6^2C_2^2r_n^{2m_0}\mathbb{E}\Big[A_{13}A_{14}A_{26}A_{63}A_{27}A_{74}\Big]\\ \nonumber
&\leq&6^2C_2^2r_n^{2m_0}\mathbb{E}\Big[A_{13}A_{14}A_{26}A_{63}A_{27}\Big]\\ \label{egsquper7}
    &=&O(r_n^{7m_0}).
\end{eqnarray}
Similarly, we have

\begin{eqnarray} \label{egsquper8}
\mathbb{E}\Big[6^2C_2^2r_n^{2m_0}A_{13} F_{14}A_{23}F_{24}\Big]&=&O(r_n^{7m_0}),\\ \label{egsquper9}
\mathbb{E}\Big[6^2C_2^2r_n^{2m_0}A_{13} F_{14}F_{23}A_{24}\Big]&=&O(r_n^{7m_0}),\\ \label{egsquper10}
\mathbb{E}\Big[6^2C_2^2r_n^{2m_0}F_{13} A_{14}A_{23}F_{24}\Big]&=&O(r_n^{7m_0}),\\ \label{egsquper11}
\mathbb{E}\Big[6^2C_2^2r_n^{2m_0}F_{13} A_{14}F_{23}A_{24}\Big]&=&O(r_n^{7m_0}),\\ \label{egsquper12}
\mathbb{E}\Big[6^2C_2^2r_n^{2m_0}F_{13} F_{14}A_{23}A_{24}\Big]&=&O(r_n^{7m_0}).
\end{eqnarray}

\begin{eqnarray}\nonumber
&&\mathbb{E}\Big[6C_2^3r_n^{3m_0}A_{13} A_{14}A_{23}F_{24}\Big]\\ \nonumber
    &=&6^2C_2^2r_n^{2m_0}\mathbb{E}\Bigg[\mathbb{E}\Big[A_{13}A_{14}A_{23}A_{27}A_{74}\Big|X_1,X_2,X_3,X_4\Big]\Bigg]\\ \nonumber
&=&6C_2^3r_n^{3m_0}\mathbb{E}\Big[A_{13}A_{14}A_{23}A_{27}A_{74}\Big]\\
&\leq&6C_2^3r_n^{3m_0}\mathbb{E}\Big[A_{13}A_{14}A_{23}A_{27}\Big]\\ \label{egsquper13}
    &=&O(r_n^{7m_0}).
\end{eqnarray}

Similarly,

\begin{eqnarray}\label{egsquper14}
\mathbb{E}\Big[6C_2^3r_n^{3m_0}A_{13} A_{14}F_{23}A_{24}\Big]=O(r_n^{7m_0}),\\ \label{egsquper15}
\mathbb{E}\Big[6C_2^3r_n^{3m_0}A_{13} F_{14}A_{23}A_{24}\Big]=O(r_n^{7m_0}),\\ \label{egsquper16}
\mathbb{E}\Big[6C_2^3r_n^{3m_0}F_{13} A_{14}A_{23}A_{24}\Big]=O(r_n^{7m_0}).
\end{eqnarray}

By (\ref{egsquper})-(\ref{egsquper16}), we get
\begin{eqnarray*}
   \mathbb{E}[G(X_3,X_4)^2]=O(r_n^{7m_0}).
\end{eqnarray*}
By assumption, $r_n=o(1)$. By (\ref{varorderder}), $\sigma_{n2}^2=\Theta(r_n^{3m_0})$. Then
\begin{eqnarray}\label{egsquper20}
  \frac{ \mathbb{E}[G(X_3,X_4)^2]}{\sigma_{n2}^2}=O\left(\frac{r_n^{7m_0}}{r_n^{6m_0}}\right)=O(r_n^{m_0})=o(1).
\end{eqnarray}

By (\ref{lemcon1}), (\ref{eg4mo}) and (\ref{egsquper20}), the assumptions of  Lemma \ref{lem4} hold. Then we have
      \begin{equation}\label{assumpn}
\frac{\sqrt{2}D_n}{n^2\sigma_{n2}}\Rightarrow N(0,1).
        \end{equation}

To finish the proof, we need to prove $\hat{\sigma}_{n2}^2=\sigma_{n2}^2(1+o_P(1))$. Recall that
\begin{eqnarray*}
    \sigma_{n2}^2=\mathbb{E}[g(X_1,X_2)^2]=\mathbb{E}[h(X_1,X_2,X_3)h(X_1,X_2,X_4)],
\end{eqnarray*}
where
\begin{eqnarray} \nonumber
h(X_1,X_2,X_3)&=&6A_{12}A_{23}A_{31}-2\left(\frac{3}{4}\right)^{m_0}\Big(A_{12}A_{23}+A_{12}A_{31}+A_{23}A_{31}\Big),
\end{eqnarray}
and
\begin{eqnarray} \nonumber
h(X_1,X_2,X_4)&=&6A_{12}A_{24}A_{41}-2\left(\frac{3}{4}\right)^{m_0}\Big(A_{12}A_{24}+A_{12}A_{41}+A_{24}A_{41}\Big).
\end{eqnarray}
Note that
\begin{eqnarray*} \nonumber
&&h(X_1,X_2,X_3)h(X_1,X_2,X_4)\\
&=&36A_{12}A_{23}A_{31}A_{24}A_{41}\\
&&-12\left(\frac{3}{4}\right)^{m_0}\Big(A_{12}A_{23}A_{31}A_{24}+A_{12}A_{23}A_{31}A_{14}+A_{12}A_{23}A_{31}A_{24}A_{41}\Big)\\
&&-12\left(\frac{3}{4}\right)^{m_0}\Big(A_{12}A_{24}A_{41}A_{23}+A_{12}A_{24}A_{41}A_{13}+A_{12}A_{24}A_{41}A_{23}A_{31}\Big)\\
&&+4\left(\frac{3}{4}\right)^{2m_0}\Big(A_{12}A_{23}A_{24}+A_{12}A_{23}A_{41}+A_{12}A_{23}A_{41}A_{42}\\
&&+A_{12}A_{13}A_{24}+A_{12}A_{13}A_{14}+A_{12}A_{13}A_{14}A_{24}\\
&&+A_{12}A_{13}A_{23}A_{24}+A_{12}A_{13}A_{23}A_{14}+A_{32}A_{31}A_{14}A_{24}\Big).
\end{eqnarray*}
Therefore,
\begin{eqnarray} \nonumber
\sigma_{n2}^2&=&\mathbb{E}[h(X_1,X_2,X_3)h(X_1,X_2,X_4)]\\ \nonumber
&=&36\mathbb{E}[A_{12}A_{23}A_{31}A_{24}A_{41}]-48\left(\frac{3}{4}\right)^{m_0}\mathbb{E}[A_{12}A_{23}A_{31}A_{24}]\\ \nonumber
&&-24\left(\frac{3}{4}\right)^{m_0}\mathbb{E}[A_{12}A_{24}A_{41}A_{23}A_{31}]+8\left(\frac{3}{4}\right)^{2m_0}\mathbb{E}[A_{12}A_{23}A_{24}]\\ \nonumber
&&+16\left(\frac{3}{4}\right)^{2m_0}\mathbb{E}[A_{12}A_{23}A_{13}A_{24}]+4\left(\frac{3}{4}\right)^{2m_0}\mathbb{E}[A_{32}A_{31}A_{14}A_{24}]\\ \nonumber
&&+8\left(\frac{3}{4}\right)^{2m_0}\mathbb{E}[A_{41}A_{12}A_{23}]\\ \nonumber
&=&\left[36-24\left(\frac{3}{4}\right)^{m_0}\right]\mathbb{E}[A_{12}A_{24}A_{41}A_{23}A_{31}]+\left[16\left(\frac{3}{4}\right)^{2m_0}-48\left(\frac{3}{4}\right)^{m_0}\right]\mathbb{E}[A_{12}A_{23}A_{13}A_{24}]\\ \nonumber
&&+8\left(\frac{3}{4}\right)^{2m_0}\mathbb{E}[A_{12}A_{23}A_{24}]+4\left(\frac{3}{4}\right)^{2m_0}\mathbb{E}[A_{32}A_{31}A_{14}A_{24}]\\ \label{varasympc}
&&+8\left(\frac{3}{4}\right)^{2m_0}\mathbb{E}[A_{41}A_{12}A_{23}].
\end{eqnarray}

Next, we show the following equations
\begin{eqnarray} \label{varconeq1}
\frac{1}{n^4}\sum_{i\neq j\neq k\neq l}A_{ij}A_{jk}A_{kl}A_{li}A_{ik}&=&\mathbb{E}[A_{12}A_{24}A_{41}A_{23}A_{31}](1+o_P(1)),\\ \label{varconeq2}
\frac{1}{n^4}\sum_{i\neq j\neq k\neq l}A_{ij}A_{jk}A_{ki}A_{il}&=&\mathbb{E}[A_{12}A_{23}A_{31}A_{24}](1+o_P(1)),\\
\label{varconeq3}
\frac{1}{n^4}\sum_{i\neq j\neq k\neq l}A_{ij}A_{ik}A_{il}&=&\mathbb{E}[A_{21}A_{23}A_{24}](1+o_P(1)),\\  \label{varconeq4}
\frac{1}{n^4}\sum_{i\neq j\neq k\neq l}A_{ij}A_{jk}A_{kl}A_{li}&=&\mathbb{E}[A_{12}A_{23}A_{34}A_{41}](1+o_P(1))\\ \label{varconeq5}
\frac{1}{n^4}\sum_{i\neq j\neq k\neq l}A_{ij}A_{jk}A_{kl} &=&\mathbb{E}[A_{41}A_{12}A_{23}](1+o_P(1)).
\end{eqnarray}

Firstly, we consider (\ref{varconeq1}). Note that
\begin{eqnarray}  \nonumber
&&\mathbb{E}\left[\left(\frac{1}{n^4}\sum_{i\neq j\neq k\neq l}\big(A_{ij}A_{jk}A_{kl}A_{li}A_{ik}-\mathbb{E}[A_{ij}A_{jk}A_{kl}A_{li}A_{ik}]\big)\right)^2\right]\\  \nonumber
&=&\frac{1}{n^8}\sum_{\substack{i\neq j\neq k\neq l\\ i_1\neq j_1\neq k_1\neq l_1}}\mathbb{E}\Big[\big(A_{ij}A_{jk}A_{kl}A_{li}A_{ik}-\mathbb{E}[A_{12}A_{24}A_{41}A_{23}A_{31}]\big)\\ \label{78eq1}
&&\times\big(A_{i_1j_1}A_{j_1k_1}A_{k_1l_1}A_{l_1i_1}A_{i_1k_1}-\mathbb{E}[A_{i_1j_1}A_{j_1k_1}A_{k_1l_1}A_{l_1i_1}A_{i_1k_1}]\big)\Big].
\end{eqnarray}

If $\{i,j,k,l\}\cap \{i_1,j_1,k_1,l_1\}=\emptyset$, then $A_{ij}A_{jk}A_{kl}A_{li}A_{ik}$ and $A_{i_1j_1}A_{j_1k_1}A_{k_1l_1}A_{l_1i_1}A_{i_1k_1}$ are independent. In this case, the expectation is zero. Hence, $\{i,j,k,l\}\cap \{i_1,j_1,k_1,l_1\}\neq\emptyset$. In this case, $|\{i,j,k,l,i_1,j_1,k_1,l_1\}|\leq 7$. 

Let $|\{i,j,k,l,i_1,j_1,k_1,l_1\}|= 7$. There are two different scenarios: $i=i_1$ and $i=l_1$. Suppose $i=i_1$. Then
\begin{eqnarray} \nonumber
&&\mathbb{E}\Big[\big(A_{ij}A_{jk}A_{kl}A_{li}A_{ik}-\mathbb{E}[A_{12}A_{24}A_{41}A_{23}A_{31}]\big)\\ \nonumber
&&\times\big(A_{i_1j_1}A_{j_1k_1}A_{k_1l_1}A_{l_1i_1}A_{i_1k_1}-\mathbb{E}[A_{i_1j_1}A_{j_1k_1}A_{k_1l_1}A_{l_1i_1}A_{i_1k_1}]\big)\Big]\\ \nonumber
&\leq&\mathbb{E}\Big[\big(A_{ij}A_{jk}A_{kl}A_{li}A_{ik}A_{ij_1}A_{j_1k_1}A_{k_1l_1}A_{l_1i_1}A_{i_1k_1}\Big]\\ \nonumber
&\leq&\mathbb{E}\Big[\big(A_{ij}A_{jk}A_{kl}A_{ij_1}A_{j_1k_1}A_{k_1l_1}\Big]\\  \label{78eq2}
&=&O(r_n^{6m_0}).
\end{eqnarray}
Here, the last equality is obtained by a similar argument as in (\ref{lempropeqp}).
The case $i=l_1$ can be similarly bounded.

Let $|\{i,j,k,l,i_1,j_1,k_1,l_1\}|= 6$. There are six different scenarios: (a). $i_1,j_1\in\{i,j,k,l\}$, (b).$i_1,k_1\in\{i,j,k,l\}$, (c). $i_1,l_1\in\{i,j,k,l\}$, (d).$j_1,k_1\in\{i,j,k,l\}$, (e).$j_1,l_1\in\{i,j,k,l\}$ and (f).$l_1,k_1\in\{i,j,k,l\}$. Suppose (a) $i_1,j_1\in\{i,j,k,l\}$. Then $k_1,l_1\in\{i,j,k,l\}$. Hence,
\begin{eqnarray}  \nonumber
&&\mathbb{E}\Big[\big(A_{ij}A_{jk}A_{kl}A_{li}A_{ik}-\mathbb{E}[A_{ij}A_{jk}A_{kl}A_{li}A_{ik}]\big)\\ \nonumber
&&\times\big(A_{i_1j_1}A_{j_1k_1}A_{k_1l_1}A_{l_1i_1}A_{i_1k_1}-\mathbb{E}[A_{i_1j_1}A_{j_1k_1}A_{k_1l_1}A_{l_1i_1}A_{i_1k_1}]\big)\Big]\\ \nonumber
&\leq&\mathbb{E}\Big[\big(A_{ij}A_{jk}A_{kl}A_{li}A_{ik}A_{ij_1}A_{j_1k_1}A_{k_1l_1}A_{l_1i_1}A_{i_1k_1}\Big]\\ \nonumber
&\leq&\mathbb{E}\Big[\big(A_{ij}A_{ik}A_{jl}A_{k_1l_1}\Big]\\   \label{78eq3}
&=&O(r_n^{4m_0}).
\end{eqnarray}
The case (b), (e) and (f) can be similarly bounded.

Suppose (c) $i_1,l_1\in\{i,j,k,l\}$. Then $j_1,k_1\in\{i,j,k,l\}$. Hence,
\begin{eqnarray}\nonumber 
&&\mathbb{E}\Big[\big(A_{ij}A_{jk}A_{kl}A_{li}A_{ik}-\mathbb{E}[A_{ij}A_{jk}A_{kl}A_{li}A_{ik}]\big)\\ \nonumber
&&\times\big(A_{i_1j_1}A_{j_1k_1}A_{k_1l_1}A_{l_1i_1}A_{i_1k_1}-\mathbb{E}[A_{i_1j_1}A_{j_1k_1}A_{k_1l_1}A_{l_1i_1}A_{i_1k_1}]\big)\Big]\\ \nonumber
&\leq&\mathbb{E}\Big[\big(A_{ij}A_{jk}A_{kl}A_{li}A_{ik}A_{ij_1}A_{j_1k_1}A_{k_1l_1}A_{l_1i_1}A_{i_1k_1}\Big]\\ \nonumber
&\leq&\mathbb{E}\Big[\big(A_{ij}A_{ik}A_{jl}A_{j_1k_1}\Big]\\ \label{78eq4}
&=&O(r_n^{4m_0}).
\end{eqnarray}

Suppose (d) $j_1,k_1\in\{i,j,k,l\}$. Then $i_1\not\in\{i,j,k,l\}$. Hence,
\begin{eqnarray} \nonumber
&&\mathbb{E}\Big[\big(A_{ij}A_{jk}A_{kl}A_{li}A_{ik}-\mathbb{E}[A_{ij}A_{jk}A_{kl}A_{li}A_{ik}]\big)\\ \nonumber
&&\times\big(A_{i_1j_1}A_{j_1k_1}A_{k_1l_1}A_{l_1i_1}A_{i_1k_1}-\mathbb{E}[A_{i_1j_1}A_{j_1k_1}A_{k_1l_1}A_{l_1i_1}A_{i_1k_1}]\big)\Big]\\ \nonumber
&\leq&\mathbb{E}\Big[\big(A_{ij}A_{jk}A_{kl}A_{li}A_{ik}A_{ij_1}A_{j_1k_1}A_{k_1l_1}A_{l_1i_1}A_{i_1k_1}\Big]\\ \nonumber
&\leq&\mathbb{E}\Big[\big(A_{ij}A_{ik}A_{jl}A_{i_1j_1}\Big]\\ \label{78eq5}
&=&O(r_n^{4m_0}).
\end{eqnarray}

Let $|\{i,j,k,l,i_1,j_1,k_1,l_1\}|\leq5$. Then
\begin{eqnarray} \nonumber
&&\mathbb{E}\Big[\big(A_{ij}A_{jk}A_{kl}A_{li}A_{ik}-\mathbb{E}[A_{ij}A_{jk}A_{kl}A_{li}A_{ik}]\big)\\ \nonumber
&&\times\big(A_{i_1j_1}A_{j_1k_1}A_{k_1l_1}A_{l_1i_1}A_{i_1k_1}-\mathbb{E}[A_{i_1j_1}A_{j_1k_1}A_{k_1l_1}A_{l_1i_1}A_{i_1k_1}]\big)\Big]\\ \nonumber
&\leq&\mathbb{E}\Big[\big(A_{ij}A_{jk}A_{kl}A_{li}A_{ik}\Big]\\ \nonumber
&\leq&\mathbb{E}\Big[\big(A_{ij}A_{jk}A_{kl}\Big]\\ \label{78eq6}
&=&O(r_n^{3m_0}).
\end{eqnarray}

By (\ref{78eq1})-(\ref{78eq6}), and the assumption that $nr_n^{m_0}=\omega(1)$, we have
\begin{eqnarray} \nonumber
&&\mathbb{E}\left[\left(\frac{1}{n^4}\sum_{i\neq j\neq k\neq l}\big(A_{ij}A_{jk}A_{kl}A_{li}A_{ik}-\mathbb{E}[A_{ij}A_{jk}A_{kl}A_{li}A_{ik}]\big)\right)^2\right]\\ \nonumber
&=& O\left(\frac{n^7r_n^{6m_0}+n^6r_n^{4m_0}+n^5r_n^{3m_0}}{n^8}\right)\\ \label{78eq7}
&=&o(r_n^{6m_0}).
\end{eqnarray}

Then (\ref{varconeq1}) follows from  (\ref{78eq7}) and Markov's inequality. Similarly, (\ref{varconeq2})-(\ref{varconeq4}) hold. By (\ref{varasympc})-(\ref{varconeq5}),  $\hat{\sigma}_{n2}^2=\sigma_{n2}^2(1+o_P(1))$.  By (\ref{assumpn}), the proof is complete.

\qed

\subsection{Proof of Theorem \ref{thm2}}

By (\ref{dnorder}) and the fact that $\hat{\sigma}_{n2}^2=\sigma_{n2}^2(1+o_P(1))$, it is straightforward to get (\ref{powerorder}).

\qed

\end{document}